\documentclass[a4paper,nobibnotes,nofootinbib]{revtex4}
\usepackage{amssymb}
\usepackage{amsmath}
\usepackage{epsfig}

\def\beq{\begin{equation}}
\def\eeq{\end{equation}}
\def\bea{\begin{eqnarray}}
\def\eea{\end{eqnarray}}
\def\bq{\begin{quote}}
\def\eq{\end{quote}}

\def\lfrestriction#1{\lower.25ex\hbox{\Big|}_{#1}}

\begin{document}

\title{Generalised parton distributions at HERA and prospects for COMPASS}

\author{L. Schoeffel}\email{laurent.schoeffel@cea.fr}
\affiliation{ CEA/Saclay, DAPNIA/Service de physique des particules, 91191 
Gif-sur-Yvette cedex, France}

\begin{abstract}
We present  a  model of generalised parton distributions based on a 
forward ansatz in the DGLAP region. We discuss some aspects of the parametrisations,
as the dependence in $t$,
with factorised and
 non-factorised approaches,
where $t$ is the square of the four-momentum  exchanged at the hadron vertex.
The predictions of this model are then compared with DVCS cross sections from H1 and ZEUS,
and a related observable, the
 skewing factor, defined as the following ratio 
 imaginary amplitudes :  $R \equiv {Im\, {\cal  A}\,(\gamma^*+p \to \gamma +p)\lfrestriction{t=0}} /
{Im \,{\cal A}\,(\gamma^*+p \to \gamma^* +p)\lfrestriction{t=0}}$. It is an interesting
quantity including both the non-forward kinematics and the
non-diagonal effects.
Finally, we discuss the beam charge asymmetry, which  is certainly the most sensitive observable to
the different hypothesis needed in the GPDs parametrisations. We show that the
approximations done for the $t$ dependence lead to significant differences for the predictions
in the HERMES kinematic domain and prospects are given for COMPASS. 

\end{abstract}
\maketitle

\section{Introduction}
\noindent

Measurements of the deep-inelastic scattering (DIS) of leptons and nucleons, $e+p\to e+X$,
allow the extraction of Parton Distribution Functions (PDFs) which describe
the longitudinal momentum carried by the quarks, anti-quarks and gluons that
make up the fast-moving nucleons. 
These functions have been measured over a wide
kinematic range in the Bjorken scaling variable $x_{Bj}$ and
the photon virtuality $Q^2$.
While PDFs provide crucial input to
perturbative Quantum Chromodynamic (QCD) calculations of processes involving
hadrons, they do not provide a complete picture of the partonic structure of
nucleons. 
In particular, PDFs contain neither information on the
correlations between partons nor on their transverse motion,
then a vital knowledge about the three dimensional 
structure of the nucleon is lost.
Hard exclusive processes, in  which the
nucleon remains intact, have emerged in recent years as prime candidates to complement
this essentially one dimentional picture \cite{ji,rad,bernard,guichon,bernard2,belitsky,bukpaper, diehlpaper, diehlreport,mueller,guzey,strikpaper,freund1,freund2}. 
This missing
information  is then encoded in Generalised Parton
Distributions (GPDs). These functions
carry information on both the longitudinal and the
transverse distribution of partons.
The recent strong interest in GPDs was stimulated by their relation
with the spin structure of the nucleon. 
Indeed, GPDs are so far the only known means of probing 
the orbital motion of partons in the nucleon through
Ji's Sum Rule \cite{ji},
which relates unpolarised GPDs to the total angular momentum of the proton.

The simplest process sensitive to GPDs is deeply virtual
Compton scattering (DVCS) or exclusive production of real photon, $e + p \rightarrow e + \gamma + p$.
This process is of particular interest as it has both a clear
experimental signature and is calculable in perturbative QCD. Also,
it does not suffer from the uncertainties caused
by the lack of understanding of the meson wave function that plague
exclusive vector meson electroproduction.
The DVCS reaction can be regarded as the elastic scattering of the
virtual photon off the proton via a colourless exchange, producing a real photon in the final state. 
In the Bjorken scaling 
regime, 
QCD calculations assume that the exchange involves two partons, having
different longitudinal and transverse momenta, in a colourless
configuration. These unequal momenta are a consequence of the mass
difference between the incoming virtual photon and the outgoing real
photon. 
The DVCS cross section depends, therefore, on GPDs \cite{bernard, guichon,bernard2,belitsky, diehlreport,mueller,guzey,strikpaper,freund1,freund2}.

The  reaction $e + p \rightarrow e + \gamma + p$ receives contributions from both the DVCS 
process, whose origin lies in the strong interaction, and the purely 
electromagnetic Bethe-Heitler (BH) process where the photon is emitted from the positron. 
Therefore, measurements can be done either of the DVCS cross section itself or of the
interference between the DVCS and BH processes, accessible via asymmetries.
Measurements of the DVCS cross section at high energy have been
obtained by H1~\cite{h1dvcs97,h1dvcs} and ZEUS~\cite{zeusdvcs} 
and the helicity asymmetry in DVCS has been measured at lower energy 
with polarised lepton beams (and unpolarised targets) by HERMES~\cite{hermesdvcs} 
and CLAS~\cite{clasdvcs,clas2}. Recently, lepton beam charge asymmetry have been
measured by HERMES \cite{hermes2}. These last observables
are very sensitive to the shape 
of the GPDs and give a good opportunity to pin-down their  parametrisations.
It is the goal of a future program at COMPASS \cite{guichon, nicole} to realise  a
dedicated measurement and to access the beam charge asymmetry under favorable conditions.

In this paper, we focus our study on the HERA results and we show that  GPDs parametrisations  based on the model of Ref. \cite{freund1,freund2}
are well suited to describe all  measurements in this range of $x_{Bj}$ ($x_{Bj}<0.1$).
In section II, we present the model.
We show the predicted DVCS cross section compared to data in section III.
In this section, we also extract the skeewing factor from data and from the model. This factor is
an interesting observable to illustrate the data/model comparison
and the building of the skewedness. In section IV, we discuss
the case of the beam charge asymmetry measured at HERMES and we give some prospects for COMPASS.

\section{Model of generalised parton distributions}

As mentioned in the introduction,
GPDs are an extension of the well-known parton distribution functions 
(PDFs) appearing in inclusive processes such as deep inelastic 
scattering (DIS), or Drell-Yan, and encode additional information 
about the partonic structure of hadrons, above and beyond that of 
conventional PDFs. 
As such, the GPDs depend on  four variables ($X,\zeta,Q^2,t$) rather than just 
 two ($x_{Bj},Q^2$) as is the case for  PDFs. We use the notations of Ref. \cite{rad},
 where the variable $X$ is defined in the range $[0,1]$ and the  variable $\zeta$ defines the skewedness with  $\zeta \simeq x_{Bj}$.
This allows an extended mapping of the dynamical behavior of the nucleon in the two 
extra variables,  the skewedness $\zeta$, and the four-momentum squared exchanged at the hadron vertex, $t$. 
In the following, we define
GPDs at a starting scale $Q_0^2$ and their $Q^2$ evolution is
generated by perturbative QCD \cite{diehlreport,mueller,guzey,strikpaper,freund1,freund2}.
They contain, in addition to the usual PDF-type information 
residing in the so-called ``DGLAP region''  (for which the momentum fraction variable is larger than the skewedness parameter, $X>\zeta$),  
supplementary information about the distribution amplitudes of virtual 
``meson-like'' states in the nucleon in the so-called ``ERBL region''
 ($X<\zeta$) \cite{diehlreport}. 
The parametrisation of GPDs is then a complicated task as the degrees of freedom are much larger than
for PDFs.

Concerning the $X,\zeta$ dependences, we follow the model presented in Ref. \cite{freund1,freund2}, using a forward ansatz at
the initial scale $Q_0=1.3$ GeV for the DGLAP domain $[\zeta,1]$ :
\begin{equation}
H_{S} (X,\zeta) \equiv \frac{q_{S}\left(\frac{X-\zeta/2}{1-\zeta/2}\right)}{1-\zeta/2}\, ,
\label{fwd}
\end{equation}
where $q_S$ refers to a singlet forward distribution \cite{cteq6} and $H_{S} (X,\zeta)=
H_{q} (X,\zeta) + H_{\bar q} (X,\zeta)$. 
A similar relation holds for valence and gluon distributions.
This Ansatz in the DGLAP region
corresponds to a double distribution model  with an
extremal profile function allowing no additional skewedness except for
the kinematical one. 
Concerning the ERBL domain, we follow the strategy of Ref. \cite{freund1,freund2}. GPDs must
be continuous at $X=\zeta$ and verify correct symmetry relations at $X=\zeta/2$.
Then, the singlet distribution can be parametrised as
\begin{equation}
H_{S} (X,\zeta) \equiv H_{S} (\zeta,\zeta) \frac{X-\zeta/2}{\zeta/2} [1+A_S(\zeta)f(X,\zeta)] 
\label{erbl}
\end{equation}
where $f(X,\zeta)$ is an even function of the variable $X-\zeta/2$
such that $f(\zeta,\zeta)=0$ to ensure the continuity relation at $X=\zeta$.
Thus, we can write : $f(X,\zeta) \propto 1-(X-\zeta/2)^2/(\zeta/2)^2$. The parameters $A_S(\zeta)$
are determined for each value of $\zeta$ by requiring that the first momentum sum rule of the GPD is satisfied \cite{ji,rad}. 
All other GPDs are parametrised as exposed in Ref. \cite{freund1,freund2}.
The skewed QCD evolution code has been rewritten for the purpose of this paper.

\begin{figure}[!htbp]
\begin{center}
 \includegraphics[totalheight=10cm]{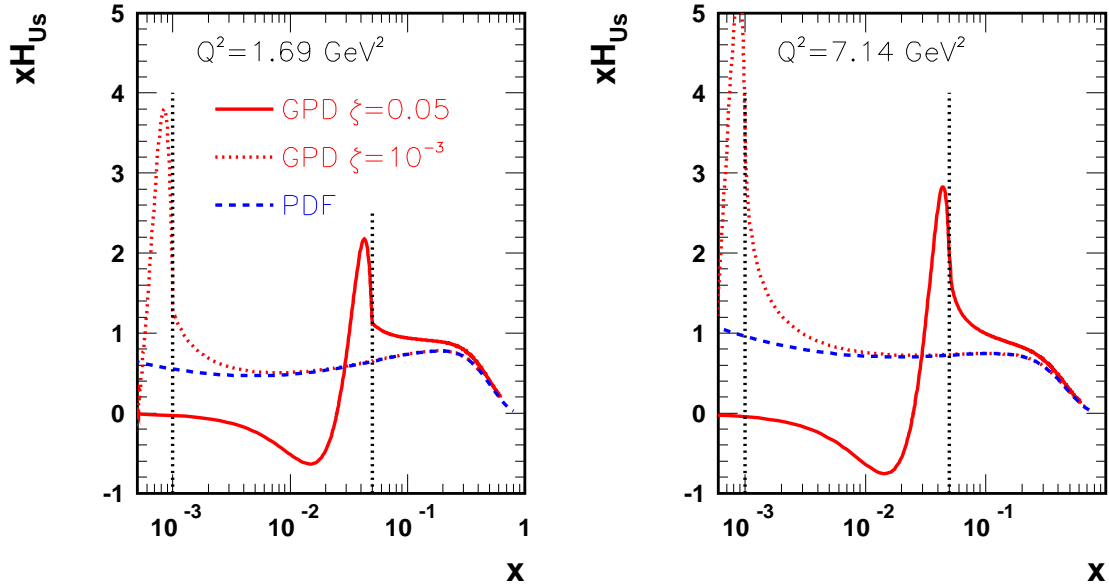}
\end{center}
\vspace*{-0.7cm}
\caption{\label{fig1}
The generalised distribution of the u quark singlet (multiplied by $X$) $X H_{Us} (X,\zeta)= X H_{u} (X,\zeta) + X H_{\bar u} (X,\zeta)$
is presented for two values of the skewedness $\zeta = 0.05$ and $10^{-3}$. The standard PDF 
$X {Us} (X)= X {u} (X) + X {\bar u} (X)$ is also displayed.
These functions are given at the initial $Q_0^2$ value of $1.69$  GeV$^2$ (left) and after the NLO QCD evolution
towards $7.14$ GeV$^2$ (right). In case of GPD, the evolution is done in the DGLAP and ERBL domain ensuring
proper continuity relations at the border \cite{freund1,freund2}.
Two vertical lines at $X=0.05$ and $X=10^{-3}$, corresponding to the skewing values, are also shown to indicate the 
position in $X$ of the transition between the DGLAP and ERBL domains.
 }
\end{figure}

\begin{figure}[!htbp]
\begin{center}
 \includegraphics[totalheight=10cm]{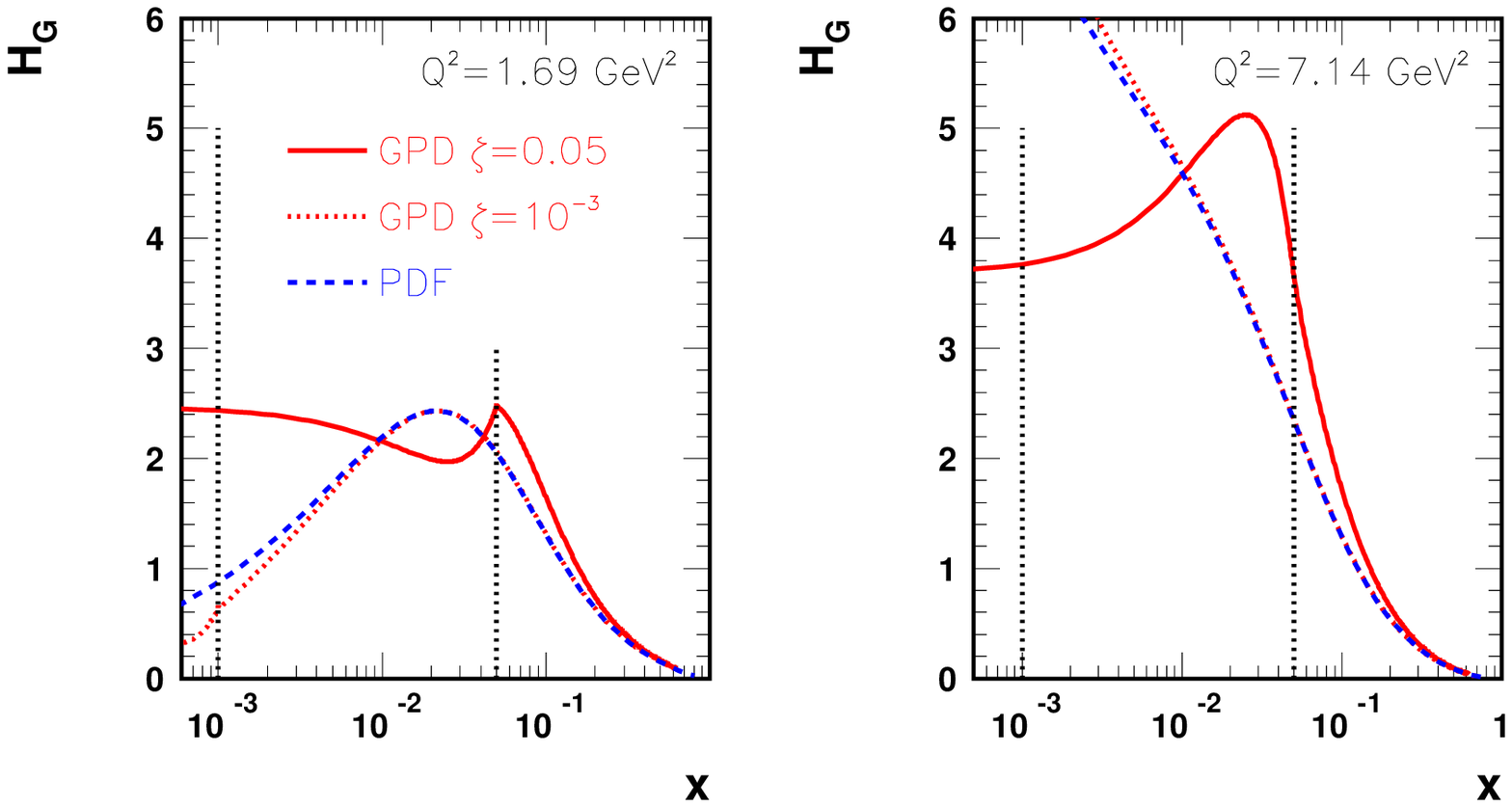}
\end{center}
\vspace*{-0.7cm}
\caption{\label{fig1b}
The generalised distribution of the gluon $H_{G} (X,\zeta)$
is presented for two values of the skewedness $\zeta = 0.05$ and $10^{-3}$. The standard PDF $XG(X)$
is also displayed.
These functions are given at the initial $Q_0^2$ value of $1.69$  GeV$^2$ (left) and after the NLO QCD evolution
towards $7.14$ GeV$^2$ (right). 
 }
\end{figure}

Results are shown in Fig. \ref{fig1} for the generalised $u$ quark  singlet density, for two values
skewedness $\zeta = 0.05$ and $\zeta =10^{-3}$, at the initial scale and after a QCD evolution.
We notice that the skewedness effect manifest itself much before the border between the DGLAP and ERBL domains at $X=\zeta$.
In particular, for the lowest $\zeta$ value of $10^{-3}$, the GPD follows exactly the PDF
till $X \simeq 10^{-2}$, where the skewing effect starts to appear, as expected from the forward limit condition : 
GPDs are obliged to reproduce the  PDFs in the forward 
limit $\zeta\to 0$, with $H_{Us}(X,\zeta=0) = Us(X) = u(X) + {\bar u}(X)$.
Also, we observe in Fig. \ref{fig1} that the generalised $u$ quark singlet distribution
is 
antisymmetric at  $X=\zeta/2$ in the ERBL region
and this property is preserved under evolution.
In Fig. \ref{fig1b}, we present the generalised gluon density again for two values
skewedness $\zeta = 0.05$ and $\zeta =10^{-3}$, at the initial scale and after a QCD evolution.
Here also, for the lowest $\zeta$ value of $10^{-3}$, the GPD follows exactly the PDF
till $X \simeq 10^{-2}$ as expected from the forward limit condition  $H_{G}(X,\zeta=0) = XG(X)$.

Concerning the $t$ dependence in the low $x_{Bj}$ kinematic domain of H1 and ZEUS measurements \cite{h1dvcs},  it has been shown that 
the
DVCS cross section, $d\sigma / dt$, can be factorised and 
approximated by an exponential form $e^{-b|t|}$,
implying a factorised dependence in $e^{-b/2 \ |t|}$ for GPDs.
It follows from Ref. \cite{h1dvcs}
that the measured $t$-slopes $b$ present a 
 $Q^2$ dependence compatible with the form :
$
b(Q^2)= A \left( 1-0.15 \log(Q^2/2) \right)
$,
where
$A=7.6 \pm 0.80$ GeV$^{-2}$. 
We keep this expression in the following when describing H1 and ZEUS measurements,
as it is done also in Ref. \cite{guzey}.
Of course, it can only be considered
as an effective approximation of a more realistic non-factorised approach \cite{diehlpaper}.
In Ref. \cite{guzey}, a non-factorised $t$ dependence is proposed with a Regge motivated approach.
In this approximation, we can write the singlet distribution at the initial scale in the DGLAP domain :
\begin{equation}
H_{S} (X,\zeta,t ; Q_0^2) \equiv  \left[ \left(\frac{1-\zeta/2}{X-\zeta/2}\right)^{\alpha'_S t }\right]  
\frac{q_{S}\left(\frac{X-\zeta/2}{1-\zeta/2}\right)}{1-\zeta/2}\, ,
\label{regge}
\end{equation}
with similar expressions for the gluon and valence distributions.
The ERBL domain is parametrised following Eq. (\ref{erbl}) with the global factor  
$\left|\frac{1-\zeta/2}{X-\zeta/2}\right|^{\alpha'_S t }$ to take into account the $t$ dependence.
The continuity at the border $X=\zeta$ is then ensured.
This parametrisation is mixing the $X$ and $t$ dependences, thus it is labeled as non-factorised. It is insprired from Regge phenomenology as
parameter $\alpha'_S$ is characteristic of a Regge trajectory. We take the values : $\alpha'_S=0.9$ GeV$^{-2}$ and $\alpha'_g=0.5$ GeV$^{-2}$
as in Ref. \cite{guzey}.

In the next sections, we compare this model with the different approximations on the $t$ dependence to 
available observables provided by H1, ZEUS and HERMES experiments.

\section{Predictions for the DVCS cross section and related observable}

\begin{figure}[!htbp]
\begin{center}
 \includegraphics[totalheight=10cm]{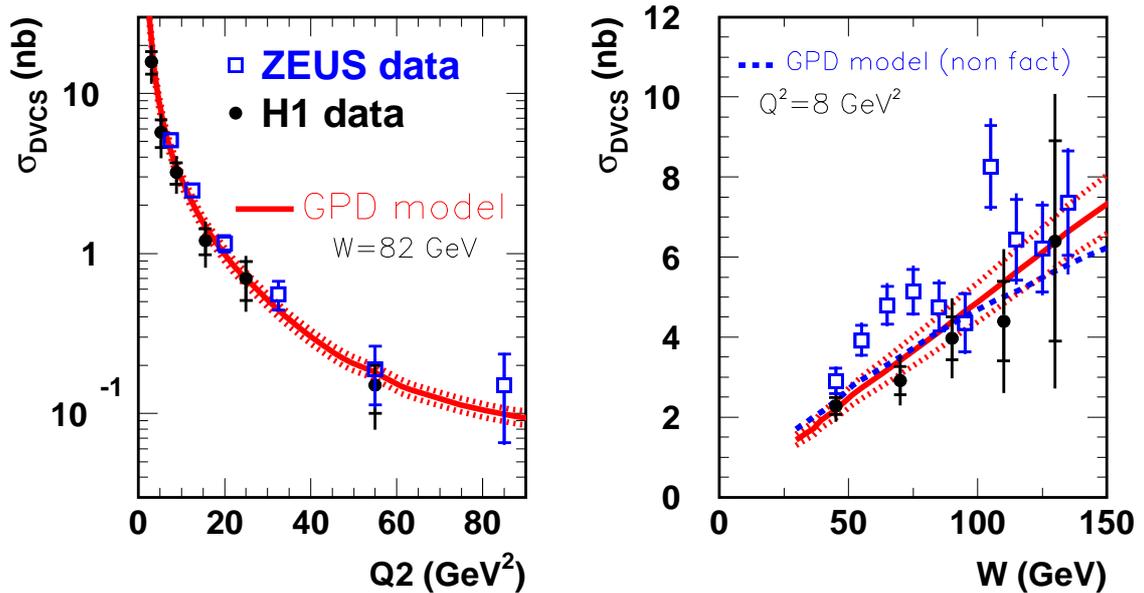}
\end{center}
\vspace*{-0.7cm}
\caption{\label{fig2}
  DVCS cross section  as a function of
  $Q^2$ at  $W=82$~GeV (left), and as a function of
$W$ at $Q^2=8$~GeV$^2$ (right). The GPD model predictions \cite{freund1,freund2} are integrated over the $t$ range of the
measurements and 
displayed (see text). 
The model of section II, using an exponential factorised $t$ dependence is shown with full lines 
on both the $Q^2$ and $W$ dependences of the DVCS cross section.
The dotted lines illustrate the uncertainty on the 
$t$-slope dependence as described in section II.
On the DVCS cross section as a function of $W$, we also display, as a dashed line, the model described in  section II using a non-factorised $t$ dependence (see text).
The results of the factorised and non-factorised approaches are close and the differences are within the dotted lines.
For clarity, we do not show the non-factorised curve on the $Q^2$ dependence plot, as it would be very close to the factorised model
on a logarithmic scale.
}
\end{figure}

Using these parametrisations of GPDs  at the initial scale $Q_0=1.3$ GeV,
calculated to higher $Q^2$ values with a skewed QCD evolution, and assumming the $t$ dependences
described above, we can provide predictions for the DVCS cross section measurements. We present the 
results in Fig. \ref{fig2} with H1 and ZEUS data \cite{h1dvcs, zeusdvcs}. 
The measurements are integrated over $t$ in the range $[-1,0]$ GeV$^2$. Then, the model predictions are
also integrated over the same range. Concerning these predictions, 
the uncertainty due to the limited knowledge of the  $t$-slopes is displayed in Fig. \ref{fig2}.
The calculations in the non-factorised $t$ approach fall within this error, which shows that the DVCS cross sections
are not sufficient to discriminate between these two hypothesis.
The same conclusion holds if we consider the 
DVCS cross section differential in $t$, $d\sigma / dt$, measured in Ref. \cite{h1dvcs}. Indeed, the 
DVCS cross sections are senstive essentially to the values of the GPDs at $X=\zeta \simeq x_{Bj}$ (at low $x_{Bj}<0.01$
for H1 and ZEUS kinematic range).
Thus, the
global factor of the non factorised approach $\left(\frac{1-\zeta/2}{X-\zeta/2}\right)^{\alpha'_S t }$ 
can be written as $\exp(-\alpha' \ln(10^{-3}) |t|)$ for H1 kinematics \cite{h1dvcs}, which is reproducing 
to a very good approximation the factorised dependence in $\exp(-b |t|)$ with $b \simeq 6$ GeV$^{-2}$.

Also, a  good agreement between data and the model predictions is observed :
it  shows that the simple description of the initial condition using a forward ansatz in the DGLAP domain and then generating
the higher $Q^2$ values from a skewed QCD evolution is a good approach. In particular, it can be noticed in Fig. \ref{fig2} 
that the $Q^2$ dependence of the DVCS cross section is well reproduced.
The $W$ dependence can be fitted by a form in $W^\delta$ with $\delta \simeq 0.7$, a large exponent value characteristic of 
a hard QCD process \cite{h1dvcs, zeusdvcs}.

It is interesting to present the DVCS cross section measurements in a form which illustrates
directly the skewing properties  by
 defining a  quantity which includes both the non-forward kinematics and the
non-diagonal effects. Namely, we set the ratio between the imaginary parts
of the DIS and DVCS (forward) scattering amplitudes at zero momentum
transfer \cite{strikman2}:
\beq
R \equiv \frac{Im\, {\cal  A}\,(\gamma^*+p \to \gamma +p)\lfrestriction{t=0}}
{Im \,{\cal A}\,(\gamma^*+p \to \gamma^* +p)\lfrestriction{t=0}} \ .
\label{R_def}
\eeq
The virtual photon is assumed to be mainly transversly polarised in the case of DVCS process
due to the real photon in the final state and has to be taken transversly polarised in the DIS
amplitude also.
The scattering amplitude for the DIS process can be  obtained from the
(transverse) DIS cross section \cite{h1f297data}, that is 
$ Im\, {\cal A}(\gamma^* p \rightarrow \gamma^* p) \sim 
\sigma_{T}(\gamma^* p \rightarrow X)$. 
In fact, the DIS amplitude can be written  using the
usual pQCD fits for the transverse proton structure function, $\sigma_{T}^{\gamma^*
p}=(4\pi^2\alpha_{EM}/Q^2)\,F_{T}^p(x,Q^2)$ and the DVCS scattering amplitude can
be obtained from the recent measurements on the Deeply Virtual Compton
Scattering cross section. 
We have shown that the DVCS cross sections are not sensitive to the hypothesis done
on the $t$ dependence described in section II. 
Then, obviously, the same conclusion holds for the skeewing factor and
 we keep, for its determination, the effective approximation supported by the H1 data \cite{h1dvcs} that
the $t$ dependence of the
amplitude can be factorised out and parameterised as
an exponential. The total DVCS cross section can
be related to the corresponding amplitude at $t=0$ :
\begin{eqnarray}
 \sigma(\gamma^*\,p\rightarrow \gamma \,p) 
 & = & \frac{\left[Im \,{\cal A}\,(\gamma^*p \to \gamma 
   p)\lfrestriction{t=0}\right]^2 }{16\pi\,b(Q^2)}\,, \label{dvcs_xs}
\end{eqnarray} 
where $b(Q^2)$ is the $t$-slope function of $Q^2$ described in section II.
The expression above is
corrected by taking into account the contribution from the real part of
the amplitude by multiplying Eq. (\ref{dvcs_xs}) by a factor $(1+\rho^2)$,
where $\rho$ is the ratio of the real to imaginary parts of the DVCS amplitude.

Following Ref. \cite{diehlivanov},
a discussion concerning the calculation of the real part in the 
forward model is necessary.
At the initial scale $Q_0=1.3$ GeV, the input GPD parametrisations in this model
do not fulfill all polynomiality relations \cite{diehlreport,diehlivanov} as only
the first momentum sum rule of the GPD is imposed  
 (see section II and Ref. \cite{freund1,freund2}\footnote{A general procedure
is also described in Ref.  \cite{freund1,freund2} to correct the input distributions
in order to verify the polynomiality relations at all orders. In the future a new version 
of the model will be proposed applying these ideas.}).
In Ref. \cite{diehlivanov},
a discussion of a simplified version of the forward model at the initial scale $Q_0$ is presented, which shows that
the calculation of the real part of the DVCS amplitude is not correct at this scale as it does not verify
the dispersion relations. The mismatch is arising mainly from a lack of flexibility of
the skewedness dependence at this very low scale. However, as soon as we let a sufficient
range in $Q^2$ for the QCD evolution to take place, the skewedness dependence is
determined mainly from the QCD evolution itself \cite{freund1,freund2}. In simple words,
it means that the memory of the original input dependence in the skewing variable 
is vanishing with the QCD evolution in $Q^2$ \cite{golec}. In the range of H1 and ZEUS data \cite{h1dvcs, zeusdvcs}, for
$Q^2 > 4$ GeV$^2$, we have
checked that the real part amplitude is compatible with the dispersion relations
to a good approximation, better that 10 \%. Then, we do not face the problem
mentioned in Ref. \cite{diehlivanov}.
Namely, the ratio of real to the imaginary parts of the DVCS amplitude, $\rho$, can
be calculated from 
the amplitudes determined in the model or using
dispersion relations. In this last case, we can write $\rho \simeq \tan (\pi\lambda
/2)$, where $\lambda=\lambda (Q^2)$ is the effective power of the
Bjorken $x_{Bj}$ dependence of the imaginary part of the amplitude.  
Hence, in the range of H1 and ZEUS data, for $Q^2 > 4$ GeV$^2$, 
we use this property to correct the skewing factor extracted from the data. In this case,
we estimate of the real part
contribution by using the effective power for the inclusive deep inelastic
reaction taken from Ref. \cite{h1lambda}. 
For the theory prediction, we use the real part of the DVCS amplitude as derived following
the model of section II. Finally, for the kinematic window available at
HERA,
the typical contribution of the term in $\rho^2$ to Eq. (\ref{dvcs_xs})   is of the order of 10 \%.

\begin{figure}[!htbp]
\begin{center}
 \includegraphics[totalheight=10cm]{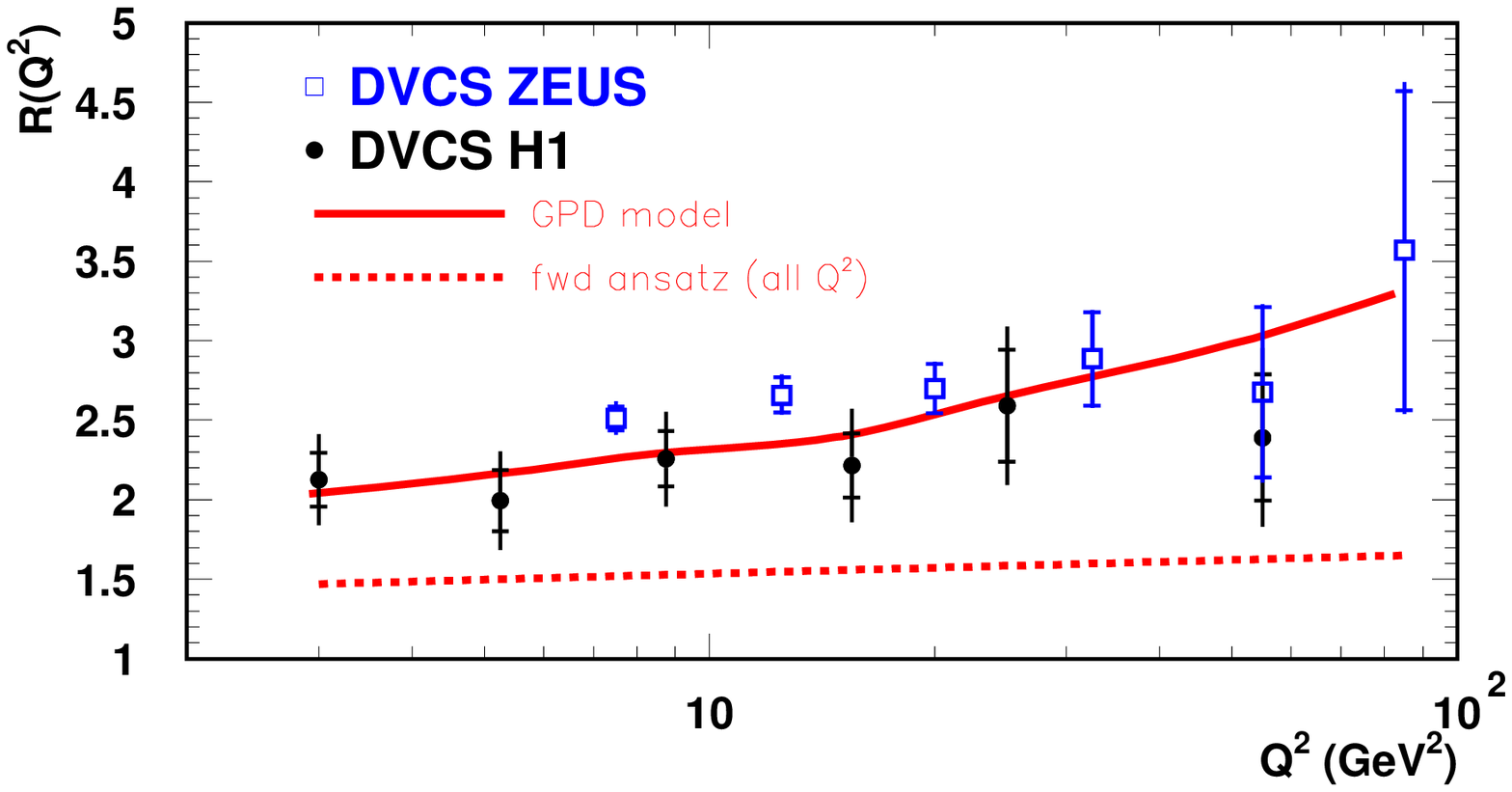}
\end{center}
\vspace*{-1.7cm}
\caption{\label{fig3}
Skewing factor
$R \equiv {Im\, {\cal  A}\,(\gamma^*+p \to \gamma +p)\lfrestriction{t=0}} /
{Im \,{\cal A}\,(\gamma^*+p \to \gamma^* +p)\lfrestriction{t=0}} $
extracted from DVCS and DIS cross sections as explained in section III.
The GPD model is also displayed and gives a good agreement of the data (full line).
The forward ansatz model, used
at all values of $Q^2$, fails to reproduce the total skewing effects generated by
the QCD evolution  (dashed line) -see text-. 
 }
\end{figure}

Considering the calculation discussed above, we can rewrite the skewing
factor as a function of the  cross sections for DIS ($\sigma_T$) and DVCS :
\begin{equation}
 R =\frac
 {4\,\sqrt{\pi \ \sigma_{DVCS}  \ b(Q^2)}}
 {\sigma_T(\gamma^* \,p\rightarrow X)\, \sqrt{(1+\rho^2)}} \,
= \frac
 {\sqrt{\sigma_{DVCS} \ Q^4 \ b(Q^2)}}
 {\sqrt{\pi^3} \,\alpha_{EM} F_T(x,Q^2)\, \sqrt{(1+\rho^2)}} \   , 
\label{R_def_ap} 
\end{equation}
Main theoretical uncertainties come from the $t$-slope, $b(Q^2)$, given
in section II. Results are shown in Fig. \ref{fig3}, where a good data/model
agreement is observed within errors.

On the skewing factor,
we can
 exemplify the part of the skewing arising from the kinematic of the DVCS process and from the $Q^2$ evolution itself. Then,
we apply the forward ansatz,
 used at the initial scale $1.3$ GeV in the model described in section II,
at all values of $Q^2$.
It means that we impose the parametrisation $
H_{S} (X,\zeta ; Q^2) \equiv \frac{q_{S}\left(\frac{X-\zeta/2}{1-\zeta/2} ; Q^2 \right)}{1-\zeta/2}$
in the DGLAP domain 
for all values of $Q^2$, and similar relations for valence and gluon distributions.
As illustrated in figure
~\ref{fig3}, the  measurements show that
such an approximation, which only takes into account the kinematical skewedness, is not sufficient to reproduce the total skewing effects generated by
the QCD evolution equations. 



\section{Beam charge asymmetry (BCA) at HERMES and COMPASS}

The determination of a cross section asymmetry with respect to the beam
charge, $(d\sigma^+ -d\sigma^-)/ (d\sigma^+ + d\sigma^-)$, has been realised 
by the HERMES experiment \cite{hermes2} for $x_{Bj} \simeq 0.1$,  $Q^2 \simeq 3$ GeV$^2$
and $|t| < 0.7$ GeV$^2$.
 The interest of this measurement lies in its large sentivity the shape of the GPD in $X,\zeta$ \cite{freund1,freund2}
 and also in the correlations between longitudinal and transverse variables \cite{guzey}.
 
 Using DVCS cross section measurements in the low $x_{Bj}$
 kinematic domain, 
 we have tested with success the GPD model described in section II. 
 In this section, we compare the predictions of this model with the results of HERMES \cite{hermes2},
 which provides measurements of the $\cos(\phi)$ amplitude of the BCA, where $\phi$ is
 the angle between the plane containing the incoming and outgoing leptons 
 and the plane defined by the virtual and real photon.
 In Table \ref{table1}, we compare the experimental values to the predictions of the 
 GPD model described in this paper,  in case of the factorised exponential $t$ dependence and 
in the the Regge non-factorised $t$  behaviour.
A table is a good way to present these results as for each $t$ values, the $x$ and $Q^2$ values
are different. 
As mentioned in the previous section, we need to let a sufficient range in $Q^2$ between the 
initial scale of the model ($Q_0=1.3$ GeV) and the measured values for the calculations of the real part of the DVCS amplitude  
to be completly correct.
We have already discussed that for $Q^2 > 4$ GeV$^2$, the exact calculation of real part of the DVCS amplitude 
in the forward model and the derivation through the dispersion relation are well compatible. In the HERMES domain,
for $Q^2 \simeq 2.5$ GeV$^2$, the agreement between both methods is of about 15 \% and less good at lower $Q^2$.
Then, we consider only the HERMES measurements with $Q^2 \ge 2.5$ GeV$^2$.

Contrary to the DVCS cross section predictions, the calculations presented in Table \ref{table1}
are very different in both approaches, factorised and non-factorised $t$ dependences.
Therefore,  we confirm that BCA  is  a very sensitive and discriminating observable to
study  GPD models. Within the present large experimental errors, as shown in Table \ref{table1},
we can not favor one approach but it is clear that further high precision measurements at COMPASS 
would be of high significance \cite{nicole}.

\begin{figure}[!htbp]
\begin{center}
 \includegraphics[totalheight=8cm]{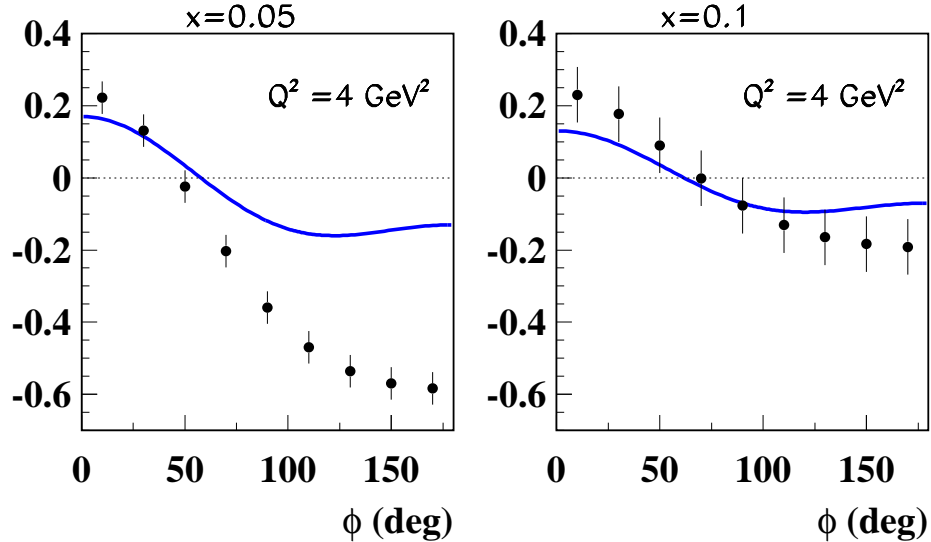}
\end{center}
\vspace*{-0.4cm}
\caption{\label{lolo}
Simulation of the azimuthal angular distribution of the beam charge
asymmetry measurable at COMPASS at $E_\mu=100$ GeV.
We present the projected values and error bars in the range
$|t|<$ 0.6 GeV$^2$
for 2 values of $x_{Bj}$ ($0.05$ and $0.1$) at $Q^2=4$ GeV$^2$ (see Ref. \cite{nicole}).
The prediction of the GPD model with a non-factorised $t$ dependence 
is shown (full line). The case of a factorised $t$ dependence would lead to
a prediction of the BCA compatible with zero and is not displayed.
 }
\end{figure}

In Fig. \ref{lolo}, we compare predictions of our model to simulations of 
the BCA extraction at COMPASS using a muon beam
of 100 GeV \cite{guichon,nicole}. We present the comparison for one value of $Q^2$ ($4$ GeV$^2$) and two values of
$x_{Bj}$ ($0.05$ and $0.1$), as we know that for $Q^2 > 4$ GeV$^2$, we stand in a safe kinematic domain
concerning the real part calculations in the forward model (see above).
When we compute the BCA in the factorised exponential $t$ dependence approximation,
we find  values compatible with zero, which are not represented in Fig. \ref{lolo}.
We just display the predictions of the model obtained in the non-factorised case.
Then, both the $\cos(\phi)$ and $\cos(2\phi)$ terms contribute to a significant level to the BCA at COMPASS,
as illustrated in Fig. \ref{lolo}.
We remark that our predictions do not match with the COMPASS simulation done with the model
of Ref. \cite{guichon}, which is another illustration of the large discriminative power of this observable
on GPDs parametrisations.

\begin{table}
\begin{center}
\begin{tabular} {|c||c|c|c||r|c|c|} 
\hline
$-t$ bin & $ \langle \, -t \, \rangle $ & $ \langle \, x_B \, \rangle $ & $ \langle \, Q^2 \, \rangle $ & \rule {0mm} {4.5mm} $A_C^{\cos \phi} \pm$ stat. $\pm$ sys. & fac. GPD & non-fac. GPD \\
(GeV$^2$) & (GeV$^2$) &         & (GeV$^2$) &  & model & model \\
\hline
\hline
0.06 -- 0.14 & 0.09 & 0.10 & 2.6 & 0.020 $\pm$ 0.054 $\pm$ 0.022  & 0.058  & 0.070 \\
\hline
0.14 -- 0.30 & 0.20 & 0.11 & 3.0 & 0.071 $\pm$ 0.066 $\pm$ 0.028  & 0.050  & 0.078 \\
\hline
0.30 -- 0.70 & 0.42 & 0.12 & 3.7 & 0.377 $\pm$ 0.110 $\pm$ 0.081 & 0.012   & 0.092 \\
\hline 
\hline 
~~~~~ $<$ 0.70 & 0.12 & 0.10 & 2.5 & 0.063 $\pm$ 0.029 $\pm$ 0.028 & 0.052   & 0.072 \\
\hline
\end{tabular}
\caption{The $\cos \phi$ amplitude of the beam--charge asymmetry per kinematic bin in
$-t$ after background correction and the respective average kinematic values.
In the last two columns, we give the predictions for the GPD model in the factorised and
non-factorised dependences of $t$.}
\label{table1}
\end{center}
\end{table}

\section{Conclusion}

An outstanding task  in QCD is related to the extraction of GPDs. 
In contrast to PDFs, these functions
contain  informations on the
correlations between partons, on their transverse motion and thus on the three dimensional 
structure of the nucleon.
Exclusive production of real photon is a prime measurement to access the GPDs,
either from DVCS cross section or from BH/DVCS interference.
Data are getting more precise, leading to more refinement in the  models.
In this paper, we have addressed the case of the kinematic domain  of HERA.
We have shown some  basic features, as skewing or $Q^2$ evolution of GPDs, in a model based on a 
forward ansatz in the DGLAP region. We have discussed the $t$ dependence in the factorised and
non-factorised approximations.
A good agreement of DVCS cross sections from H1 and ZEUS have been obtained in both cases.
We have compared this model with success also on the
 skewing factor $R \equiv {Im\, {\cal  A}\,(\gamma^*+p \to \gamma +p)\lfrestriction{t=0}} /
{Im \,{\cal A}\,(\gamma^*+p \to \gamma^* +p)\lfrestriction{t=0}}$, which is an interesting
quantity as it includes both the non-forward kinematics and the
non-diagonal effects.
Finally, we have shown that the BCA  is  an interesting sensitive observable to
the different hypothesis needed in the GPDs parametrisations. We have shown that the
approximations done for the $t$ dependence lead to significant differences for the predictions
in the HERMES kinematic domain, and even larger for COMPASS. 

\section*{Acknowledgements}

 I would like to thank N. d'Hose, P. Guichon and E. Sauvan for useful remarks  and comments on 
 the manuscript.


\end{document}